\documentclass[conference]{IEEEtran}
\usepackage{hyperref}

\IEEEoverridecommandlockouts

\usepackage{cite,amssymb}
\usepackage{color}
\usepackage{amsthm}

\usepackage{graphicx}
\graphicspath{{./Figs/}}

\usepackage[cmex10]{amsmath}
\newcommand{\R}{\mathbb{R}}

\newcommand{\E}{\operatorname{E}}

\newcommand{\e}{\mathrm{e}}

\newcommand{\vct}[1]{\boldsymbol{#1}}
\newcommand{\mtx}[1]{\boldsymbol{#1}}
\newcommand{\<}{\langle}
\renewcommand{\>}{\rangle}
\newcommand{\T}{{\scalebox{0.5}[0.6]{$\mathsf{T}$}}}

\newcommand{\trace}{\operatorname{trace}}

\newcommand{\set}[1]{\mathcal{#1}}

\newcommand{\va}{\vct{a}}

\newcommand{\ve}{\vct{e}}

\newcommand{\vw}{\vct{w}}
\newcommand{\vx}{\vct{x}}
\newcommand{\vy}{\vct{y}}
\newcommand{\vz}{\vct{z}}
\newcommand{\vzero}{\vct{0}}

\newcommand{\mB}{\mtx{B}}

\newcommand{\mQ}{\mtx{Q}}

\newcommand{\mU}{\mtx{U}}
\newcommand{\mV}{\mtx{V}}

\newcommand{\mZ}{\mtx{Z}}

\newcommand{\mPsi}{\mtx{\varPsi}}
\newcommand{\mSigma}{\mtx{\varSigma}}

\newcommand{\mId}{{\bf I}}

\newcommand{\mzero}{{\bf 0}}

\newcommand{\setT}{\set{T}}

\newcommand{\sK}{K}
\newcommand{\sL}{L}
\newcommand{\sM}{M}
\newcommand{\sN}{N}	
	
\newcommand{\sQ}{Q}
\newcommand{\sR}{R}
\newcommand{\sT}{T}

\newcommand{\linop}[1]{\mathcal{#1}}
\newcommand{\loW}{\linop{W}}

\DeclareMathOperator*{\argmin}{argmin}

\global\long\def\mb#1{\boldsymbol{#1}}

\global\long\def\mr#1{\mathrm{#1}}
\global\long\def\msf#1{\mathsf{#1}}
\theoremstyle{plain}
\newtheorem{thm}{\protect\theoremname}
\theoremstyle{plain}
\newtheorem{lem}{\protect\lemmaname}

\providecommand{\theoremname}{Theorem}
\providecommand{\lemmaname}{Lemma}

\begin{document}

\title{Sketching for Simultaneously Sparse and Low-Rank Covariance Matrices}

%\author{\IEEEauthorblockN{Michael Shell}
%\IEEEauthorblockA{School of Electrical and\\Computer Engineering\\
%Georgia Institute of Technology\\
%Atlanta, Georgia 30332--0250\\
%Email: http://www.michaelshell.org/contact.html}
%\and
%\IEEEauthorblockN{Homer Simpson}
%\IEEEauthorblockA{Twentieth Century Fox\\
%\and
%\IEEEauthorblockN{James Kirk\\ and Montgomery Scott}
%\IEEEauthorblockA{Starfleet Academy\\
%San Francisco, California 96678-2391\\

\makeatletter
\author{\IEEEauthorblockN{Sohail~Bahmani and  Justin~Romberg\thanks{This work was supported by ONR grant N00014-11-1-0459, and NSF grants CCF-1415498 and CCF-1422540.}}
\IEEEauthorblockA{School of Electrical and Computer Engineering\\
Georgia Institute of Technology\\
Atlanta, GA 30332\\
Email: \{sohail.bahmani,jrom\}@ece.gatech.edu}
}
\makeatother

%------------------------------------------------------------------------------

\maketitle

\begin{abstract}
%\boldmath
We introduce a technique for estimating a structured covariance matrix from observations of a random vector which have been sketched.  Each observed random vector $\vx_t$ is reduced to a single number by taking its inner product against one of a number of pre-selected vector $\va_\ell$.  These observations are used to form estimates of linear observations of the covariance matrix $\mSigma$, which is assumed to be simultaneously sparse and low-rank.  We show that if the sketching vectors $\va_\ell$ have a special structure, then we can use straightforward two-stage algorithm that exploits this structure.  We show that the estimate is accurate when the number of sketches is proportional to the maximum of the rank times the number of significant rows/columns of $\mSigma$.  Moreover, our algorithm takes direct advantage of the low-rank structure of $\mSigma$ by only manipulating matrices that are far smaller than the original covariance matrix.
\end{abstract}

%------------------------------------------------------------------------------
\section{Introduction}

We introduce and analyze a technique for estimating the covariance matrix $\mSigma$ of $\sN$-dimensional random vector $\vx$ from samples $\vx_1,\vx_2,\ldots,\vx_{\sQ}$.  We will show that these quantities can be estimated accurately from {\em sketches} or {\em compressed measurements} of $\vx$. Different methods for sketching covariance matrices that are either Toeplitz, sparse, or low-rank are studied in \cite{bioucas-dias14co} and \cite{dasarathy15sk}. Sketching of simultaneously structured covariance matrices (e.g., low-rank and sparse matrices or low-rank and Toeplitz matrices) was first considered in \cite{chen14es} and \cite{chen15ex}. In particular, it is shown in \cite{chen14es} that simultaneously $\sK\times \sK$-sparse and rank-$\sR$ covariance matrices can be recovered from $\msf{O}(\sK^2 \sR\log \sN)$ generic rank-one sketeches through minimizaton of a mixture of the trace norm and the $\ell_1$ norm.  It is recognized in \cite{chen14es} that the achieved sample complexity is suboptimal. In this paper, we show how the sample complexity can be improved using specifically tailored rank-one sketches. We also demonstrate how the estimation can be performed by manipulating matrices of dimension $\sN\times\sR$, where $\sR$ is the rank of the target.   

As the samples $\vx_t$ are presented, they are mapped into scalar values by taking an inner product against one of $\sL$ different vectors $\va_1,\ldots,\va_{\sL}$.  If a fixed vector $\va_\ell$ is used on $\sT$ different samples, then we have the estimate
\begin{equation}
	\label{eq:approxmean}
	\va_\ell^\T\mSigma\va_\ell = \E\left[\va_\ell^\T\vx\vx^\T\va_\ell\right] \approx \frac{1}{\sT}\sum_{t=1}^{\sT}|\<\vx_t,\va_{\ell}\>|^2.
\end{equation}
The $\sQ$ data points are thus turned into $\sL$ measurements of $\mSigma$ of the form
\begin{equation}
	\label{eq:ymeas}
	y[\ell] = \va_\ell^\T\mSigma\va_\ell + \mathrm{noise}.
\end{equation}
The measurements $y[\ell]$ can be formed in a decentralized manner, and since they are scalars they are easy to store and communicate.  We will show that by choosing the vectors $\va_\ell$ attentively, we can estimate $\mSigma$ from $\sL=\msf{O}(\sK\sR\log \sN)$ rank-one sketches that is much smaller than the number of entries in the covariance matrix, $\sL\ll\sN^2$. The proposed sketching scheme is similar to the efficient measurement scheme recently proposed for compressive phase retrieval \cite{iwen15ro,bahmani15ef}.  Moreover, our algorithm for estimating $\mSigma$ works by manipulating matrices in factored form, making it scalable for large $\sN$ regimes.

\subsection{Data model}

The data points $\vx_1,\ldots,\vx_{\sQ}\in\R^{\sN}$ are independent realizations of a zero-mean random vector with covariance matrix $\mSigma$.  We will consider the case where $\mSigma$ is simultaneously sparse and low-rank.  This means that we can closely approximate $\mSigma$ with the factorization
\[
	\mSigma \approx \mU\mU^\T,
\]
where $\mU$ is a $\sN\times\sR$ matrix with at most $\sK$ non-zero rows (we will assume $\sR\leq\sK\ll\sN$).   $\mSigma$ itself can be well-approximated by a matrix with $\sK^2$ non-zero terms --- all but $\sK$ rows (or columns) will be approximately zero, and in each of these $\sK$ significant rows, all but $\sK$ entries will be approximately zero.

Other than these properties of the covariance matrix, our framework does not depend heavily on the particulars of the distribution of the $\vx_q$.  The algorithms below depend on having bounds on the approximation error in \eqref{eq:approxmean}; these bounds might be derived from other known properties of $\mSigma$ (i.e.\ its trace or the dynamic range of its eigenvalues) or of the distribution of $\vx$.

\subsection{Choosing the sketching vectors}
\label{sec:sketchingvectors}

Each data vector $\vx_t$ is compressed into a single number by taking an inner product against one of $L$ different $\va_\ell\in\R^{\sN}$.  We will denote the set of indices that use vector $\va_\ell$ as $\setT_\ell$.  It is not critically important how the $\sQ$ observation are divided among the $\setT_\ell$ other than that the sets should have close to the same size.  For simplicity, we will assume that all of the index sets have the same number of terms, $T:=|\setT_\ell|$ for all $\ell$, and so $\sQ=\sT\sL$.

The guarantees for the estimation algorithm presented below depend on the $\va_\ell$ having certain properties.  We restrict the $\va_\ell$ to lie in an $\sM$ dimensional subspace of $\R^{\sN}$, generating them using
\begin{equation}
	\label{eq:asubspace}
	\va_{\ell} = \mPsi^\T\vw_\ell,
\end{equation}
where $\mPsi$ is an $\sM\times\sN$ matrix whose rows form a basis for this subspace.  We will take the $\vw_\ell$ to be randomly generated.  The analysis below holds when the $\vw_\ell$ are independent and distributed $\mathrm{Normal}(\vzero,\mId)$.  However, it is likely that the analysis could be generalized to other distributions, and the only thing we rely on algorithmically is that all $\sL$ vectors lie in a subspace as in \eqref{eq:asubspace}.

Our analysis also requires that the matrix $\mPsi$ is a stable embedding for $2\sK$-sparse vectors.  We will assume below that $\mPsi$ obeys the {\em restricted isometry property} \cite{candes06ne}
\begin{equation}
	\label{eq:rip}
	(1-\delta_{\sK})\|\vz_1-\vz_2\|_2^2
	~\leq~
	\|\mPsi(\vz_1-\vz_2)\|_2^2
	~\leq~
	(1+\delta_{\sK})\|\vz_1-\vz_2\|_2^2,
\end{equation}
for all $\sK$-sparse $\vz_1,\vz_2\in\R^{\sN}$.  There are many examples of matrices that have this property for a number of rows proportional to the sparsity $\sK$ times a logarithmic factor; see \cite{vershynin12in,rauhut10co} for detailed overviews.  If $\mPsi$ is populated with independent \emph{subgaussian} random variables, then \eqref{eq:rip} holds with high probability for $\sM\gtrsim\sK\log(\sN/\sK)$.  More structured matrices that allow for fast computations also have this property.  For example, if $\mPsi$ consists of $\sM$ randomly selected rows of an $\sN\times\sN$ Fourier matrix, then \eqref{eq:rip} holds with high probability for $\sM\gtrsim\sK\log^4(\sN)$. 

With this subspace conditions, we can write the covariance measurements as
\begin{align}
	\nonumber
	y[\ell] = \frac{1}{T}\sum_{t\in\setT_\ell}|\<\vx_t,\va_\ell\>|^2 &\approx \vw_\ell^\T\mPsi\mSigma\mPsi^\T\vw_\ell \\
	\label{eq:rank1ip}
	&= \<\mPsi\mSigma\mPsi^\T,\vw_\ell\vw_\ell^\T\>.
\end{align}
That is, each $y[\ell]$ is the matrix inner product of $\mPsi\mSigma\mPsi^\T$ with an $\sM\times\sM$ rank-1 random matrix.  This gives the covariance matrix measurements a {\em nested structure}.  We can write
\[
	\vy = \loW(\mPsi\mSigma\mPsi^\T) + \mathrm{noise}.
\] 
The $\sN\times\sN$ covariance matrix $\mSigma$ is first mapped to an $\sM\times\sM$ matrix by applying $\mPsi$ to either side.  Then $\loW$ maps this result to $\R^{\sL}$ by taking the series of matrix inner products in \eqref{eq:rank1ip}.  

In the next section, we will see how this nested structure allows us to decouple the estimation process into a low-rank estimation stage followed by two sparse approximation stages.

In Section~\ref{sec:analysis}, we present results that relate the number of sketches $\sL$ and their accuracy (which is controlled by $\sT$) to the accuracy of our estimation procedure.

\subsection{Estimating the covariance}
\label{sec:algorithm}

Let $\mSigma^{\star}\in\R^{\sN\times\sN}$ be a rank-$\sR$ positive semidefinite matrix that has at most $\sK$ nonzero rows (and columns) with $\sR < \sL\ll \sN$.  Given a matrix $\mPsi\in\R^{\sM\times\sN}$ and a linear operator
$\loW:\mB\mapsto\left[\vw_{\ell}^\T\mB\vw_\ell\right]_{\ell=1}^{\sL}$, where $\vw_\ell\sim\mathrm{Normal}\left(\vzero,\mId\right)$,
we consider the problem of estimating $\mSigma^{\star}$ from noisy linear measurements of the form 
\begin{align}
	\vy & =\loW\left(\mPsi\mSigma^\star\mPsi^\T\right) + \vz,
	\label{eq:measurements}
\end{align}
where $\vz\in\R^{\sL}$ is the noise vector that is bounded as $\|\vz\|_2\leq\varepsilon$. 

We propose the following two-stage procedure for estimation of $\mSigma^{\star}$. 
\begin{enumerate}

	\item {\bf Low-rank estimation stage}:  Since $\mSigma^\star$ is low-rank, we know that $\mPsi\mSigma\mPsi^\T$ will be low-rank as well.  Our first stage ``inverts'' the $\loW(\cdot)$ operator by looking for a low-rank matrix that explains the measurements $\vy$.  There are a number of ways we might do this, but here we will use the standard convex relaxation that minimizes the trace norm (nuclear norm) in place of the rank:
	\begin{align}
		\widehat{\mB} & \in\argmin_{\mB\succcurlyeq\mzero}\ \trace\left(\mB\right) 
		\label{eq:pre-estimate}\\
 		&\hspace{3ex}\mr{subject\ to\ }\|\loW\left(\mB\right)-\vy\|_{2}\leq\varepsilon.\nonumber 
	\end{align}
	The output $\widehat{\mB}$ of this first stage can be thought of as an estimate of $\mPsi\mSigma^\star\mPsi^\T$.
	
	\item {\bf Sparse estimation stage}:  In the second stage, we invert the action of $\mPsi$ on the left and $\mPsi^\T$ on the right.  It is conceivable that these two steps could be combined into a single sparse approximation step, but performing them sequentially leads to a natural analysis (as we will see in Section~\ref{sec:analysis}).

	\begin{enumerate}
	
		\item \label{enu:sparse-a}
		Since we are given $\widehat{\mB}\approx\mPsi\mSigma^\star\mPsi^\T$, we start by looking for a matrix with a small number of non-zero rows whose range is close to the range of $\mSigma^\star$.  The fact that $\widehat{\mB}$ is approximately rank $\sR$ allows us to work with a rank $\sR$ approximation of $\widehat{\mB}$ and be computationally efficient.  We start by computing the $\sM\times\sR$ matrix $\widehat{\mU}$ of the top $\sR$ (unit-norm) eigenvectors of $\widehat{\mB}$.  We can then look for a row-wise sparse $\sN\times\sR$ matrix that is close to $\widehat{\mB}\widehat{\mU}$ after an application of $\mPsi$ on the left:
		\begin{align}
			\widehat{\mQ}_{1} & \in\argmin_{\mQ}\left\|\mQ\right\|_{1,2} \nonumber\\
 			&\hspace{3ex}\mr{subject\ to\ }\left\Vert \mPsi\mQ-\widehat{\mB}\widehat{\mU}\right\Vert _{F}\leq\frac{c_{1}\varepsilon}{\sqrt{\sL}}.
 			\label{eq:sparseaopt}
		\end{align}
		The $\|\cdot\|_{1,2}$ norm above is a convex relaxation for the number of non-zero rows; $\|\mZ\|_{1,2} = \sum_{n=1}^{\sN}\|\mZ_{n,:}\|_2$, where $\mZ_{n,:}$ is the $n$th row of $\mZ$.  Conceptually (and this is formalized in the analysis below), the output $\widehat{\mQ}_1$ can be used to form the approximation $\widehat{\mQ}_1\widehat{\mU}^\T\approx\mSigma^\star\mPsi^\T$.  Thus we have effectively undone the action of $\mPsi$ on the left of $\mSigma^\star$.

		\item \label{enu:sparse-b}
		The second sparse approximation step does a similar computation on the row-space of the output above.  We
		compute $\widehat{\mV}\in\mSigma^{\sN\times\sR}$, the matrix of the left singular vectors of $\widehat{\mQ}_{1}$, and compute 
		\begin{align}
			\widehat{\mQ}_{2} & \in\argmin_{\mQ}\left\Vert \mQ\right\Vert _{1,2} \nonumber\\
 			&\hspace{3ex}\mr{subject\ to\ }\left\Vert \mPsi\mQ- \widehat{\mU}\widehat{\mQ}_{1}^\T\widehat{\mV}\right\Vert _{F}\!\!\leq\!\frac{c_{2}\varepsilon}{\sqrt{\sL}}.
 			\label{eq:sparsebopt}
		\end{align}
		Conceptually, the output $\widehat{\mQ}_{2}$ undoes the action of $\mPsi^\T$ on the right, giving us the approximation $\widehat{\mQ}_{2}\widehat{\mV}^\T\approx\mSigma^\star$
		
	\end{enumerate}

\end{enumerate}

The output of the second stage are two $\sN\times\sR$ matrices $\widehat{\mV}$ and $\widehat{\mQ}_2$.  These can be used to form an approximate factorization of $\mSigma^\star$.  In Section~\ref{sec:analysis} below, we give a mathematical guarantee on how close
\begin{align}
		\widehat{\mSigma} & =\widehat{\mQ}_{2}\widehat{\mV}^\T,
		\label{eq:estimate}
\end{align}
is to the true covariance $\mSigma^\star$.

The constants $c_1$ and $c_2$ in the sparse estimation stage above are user-defined; the analysis below relies on a particular choice for these parameters.

The algorithm above is carefully designed to be sublinear in the number of entries in the $\sN\times\sN$ matrix $\mSigma^\star$.  Stage 1 above is an optimization program over the cone of $\sM\times\sM$ SDP matrices, and we have seen that we can take $\sM\gtrsim\sK\log^\alpha\sN$.  The sparse approximation stage exclusively handles $\sN\times\sR$ matrices.  This allows our algorithmic framework to scale to regimes where the dimension $\sN$ and the number of samples $\sQ$ are large.

\subsection{Noise magnitude}
In general, all the algorithm proposed above needs is a bound on the total size of the error in the measurements of the covariance matrix.  With the model in \eqref{eq:approxmean} and \eqref{eq:ymeas}, this error is simply the deviation of a quadratic function of the random vector $\vx$ from its mean.  If we have information about the distribution of $\vx$, we may be able to derive the desired bound in a principled way.  

To demonstrate this, suppose that $\vx\sim\mr{Normal}(\mb{0},\mSigma)$ which implies $\<\vx,\va_\ell\>\sim\mr{Normal}(0,\va_\ell^\T\mSigma\va_\ell)$.  This means that 
\[
	y[\ell] = \frac{1}{\sT}\sum_{t\in\setT_\ell}|\<\vx_t,\va_\ell\>|^2
\]	
is proportional to a Chi-squared random variable with $\sT$ degrees of freedom. Since $\E[y[\ell]]=\va_\ell^\T\mSigma\va_\ell$, the mean energy of the $\ell$th component $e[\ell]:=y[\ell]-\va_\ell^\T\mSigma\va_\ell$ of the error is
\[
	\E[e[\ell]^2]  = \frac{2(\va_\ell^\T\mSigma\va_\ell)^2}{T},
\]
and the total error $\|\ve\|_2^2$ can be shown to concentrate around
\begin{align*}
	\E[\|\ve\|_2^2] &= \frac{2}{T}\sum_{\ell=1}^L(\va_\ell^\T\mSigma\va_\ell)^2.
\end{align*}
The obtained concentration bounds may depend on the (a priori unknown) covariance $\mSigma$, but usually they can be approximated by some attributes of $\mSigma$.  If the $\va_\ell$ are chosen as in Section~\ref{sec:sketchingvectors}, $\va_\ell^\T\mSigma\va_\ell$ will itself be a weighted sum of Chi-squared random variables whose moments can be calculated in terms of the Frobenius and spectral norms of $\mSigma$.
%------------------------------------------------------------------------------
\section{Analysis}
\label{sec:analysis}

Our main theorem shows that for an appropriate number of sketches, the estimation algorithm detailed in Section~\ref{sec:algorithm} produces a provably good estimate of $\mSigma^\star$.  In addition to the number of sketches $\sL$ being sufficiently large, we also assume that $\mPsi$ obeys the restricted isometry property in \eqref{eq:rip}.

\begin{thm}
	\label{thm:accuracy} 
	There is a constant $C_1$ such that if
	\begin{equation}
		\label{eq:Lbound}
		\sL ~\geq~ C_1\cdot \sR\cdot \sM,
	\end{equation}
	then for appropriately chosen constants $c_{1}$ and $c_{2}$ in \eqref{eq:sparseaopt} and \eqref{eq:sparsebopt} above, with probability exceeding $1-\e^{-\msf{O}(M)}$ the estimate in \eqref{eq:estimate} obeys
	\begin{align*}
		\left\Vert \widehat{\mSigma}-\mSigma^{\star}\right\Vert _{F} & \leq\frac{C_2\varepsilon}{\sqrt{\sL}},
	\end{align*}
	where $C_2$ is another constant which depends on $c_1,c_2,$ and the restricted isometry constant $\delta_{\sK}$.
	%for all rank-$r$ and row-wise $k$-sparse $\mb X^{\star}$ with high
	%probability.
\end{thm}

The theorem above gives us a uniform guarantee; it holds simultaneously for all rank-$\sR$ and row-wise $\sK$-sparse $\mSigma^\star$ for the same $\{\va_\ell\}$.  We will sketch a proof below, withholding some of the details due to space constraints.

The accuracy of the first stage is established through a simple application of a recent result in the theory of low-rank matrix recovery.  The work \cite{kueng14lo} establishes uniform bounds on the recovery of low-rank matrices from inner products against a series of independent, rank-1 symmetric random matrices.  This exactly describes the $\loW(\cdot)$ operator in \eqref{eq:measurements}.  A direct application of the main theorem in that paper shows that for $\sL$ as in \eqref{eq:Lbound}, we have 
\begin{alignat}{1}
	\left\Vert \widehat{\mB}-\mPsi\mSigma^{\star}\mPsi^\T\right\Vert _{F} & \leq\frac{c\varepsilon}{\sqrt{\sL}},
	\label{eq:low-rank-accuracy}
\end{alignat}
with high probability for an absolute constant $c>0$. 

The accuracy of the second stage follows from the lemma below, which we will prove at the end of the section.
\begin{lem}
	\label{lem:master}
	Suppose that for some rank-$\sR$ and row-wise $\sK$-sparse matrix $\mSigma^{\sharp}$, a matrix $\mB^{\sharp}$, and a constant $\epsilon$ we have
	\begin{align*}
		\left\Vert \mPsi\mSigma^{\sharp}-\mB^{\sharp}\right\Vert _{F} & \leq\epsilon,
	\end{align*}
 	where $\mPsi$ obeys \eqref{eq:rip} with a sufficiently small $\delta_{\sK}$. Let $\mU^{\sharp}$ denote the top
$\sR$ right singular vectors of $\mB^{\sharp}$ and 
	\begin{align}
		\mQ^{\sharp} & =\argmin_{\mQ}\ \left\Vert \mQ\right\Vert _{1,2} \label{eq:Q-opt}\\
 		&\hspace{3ex} \mr{subject\ to\ }\left\Vert \mPsi\mQ-\mB^{\sharp}\mU^{\sharp}\right\Vert _{F}\leq2\epsilon.\nonumber 
	\end{align}
 	Then we have 
	\begin{align*}
		\left\Vert \mQ^{\sharp}\mU^{\sharp\T}-\mSigma^{\sharp}\right\Vert _{F} & \leq C\epsilon,
	\end{align*}
 	for some constant $C>0$ depending only on $\delta_{\sK}$.
\end{lem}

With $\widehat{\mU}$ as the top $\sR$ eigenvectors of the output of the first stage $\widehat{\mB}$, Lemma~\ref{lem:master} tells us that the result $\widehat{\mQ}_1$ of solving \eqref{eq:sparseaopt} in Stage 2a will obey
\begin{align*}
	\left\Vert \widehat{\mU}\widehat{\mQ}_{1}^\T-\mPsi\mSigma^{\star}\right\Vert _{F}=
	\left\Vert \widehat{\mQ}_{1}\widehat{\mU}^\T-\mSigma^{\star}\mPsi^\T\right\Vert _{F} 
	& \leq\frac{c'\varepsilon}{\sqrt{\sL}},
\end{align*}
for some constant $c'$.  For stage 2b, with $\widehat{\mV}\in\R^{\sN\times\sR}$ as the top $\sR$ left singular vectors of $\widehat{\mQ}_1$, we can again invoke the lemma and conclude that for some constant $C_2$ we have
\begin{align*}
	\left\Vert \widehat{\mSigma}-\mSigma^{\star}\right\Vert _{F} 
	&=\left\Vert \widehat{\mQ}_{2}\widehat{\mV}^\T-\mSigma^{\star}\right\Vert _{F}
	\leq\frac{C_{2}\varepsilon}{\sqrt{\sL}}.
\end{align*}

It remains to prove Lemma~\ref{lem:master}.  Because $\mPsi\mSigma^{\sharp}$ is rank-$\sR$ and $\mB^{\sharp}\mU^{\sharp}\mU^{\sharp\T}$
is the best rank-$\sR$ approximation of $\mB^{\sharp}$, we have
\begin{align*}
	\left\Vert \mPsi\mSigma^{\sharp}-\mB^{\sharp}\mU^{\sharp}\mU^{\sharp\T}\right\Vert _{F} 
	& \leq\left\Vert \mPsi\mSigma^{\sharp}\!-\!\mB^{\sharp}\right\Vert _{F}\!\! +\!
	\left\Vert \mB^{\sharp}\!-\!\mB^{\sharp}\mU^{\sharp}\mU^{\sharp\T}\right\Vert _{F}\\
 	& \leq 2\left\Vert \mPsi\mSigma^{\sharp}-\mB^{\sharp}\right\Vert _{F}
 	\leq 2\epsilon.
\end{align*}
Therefore, we can write 
\begin{align*}
	&\left\Vert \mPsi\mSigma^{\sharp}\left(\mId-\mU^{\sharp}\mU^{\sharp\T}\right)\right\Vert _{F}^{2} +
	\left\Vert \left(\mPsi\mSigma^{\sharp}-\mB^{\sharp}\right)\mU^{\sharp}\right\Vert _{F}^{2} \\
	& =\left\Vert \mPsi\mSigma^{\sharp}\left(\mId - \mU^{\sharp}\mU^{\sharp\T}\right)\right\Vert _{F}^{2}
	+ \left\Vert \left(\mPsi\mSigma^{\sharp}-\mB^{\sharp}\right)\mU^{\sharp}\mU^{\sharp\T}\right\Vert _{F}^{2}\\
 	& =\left\Vert \mPsi\mSigma^{\sharp}-\mB^{\sharp}\mU^{\sharp}\mU^{\sharp\T}\right\Vert _{F}^{2}
 	\leq 4 \epsilon^{2}.
\end{align*}
In particular, since $\mSigma^\sharp$ is row-wise $\sK$-sparse, from \eqref{eq:rip} we have 
\begin{align}
	\sqrt{1-\delta_{\sK}}\left\Vert \mSigma^{\sharp}\left(\mId-\mU^{\sharp}\mU^{\sharp\T}\right)\right\Vert _{F}
	& \leq\left\Vert \mPsi\mSigma^{\sharp}\left(\mId-\mU^{\sharp}\mU^{\sharp\T}\right)\right\Vert _{F}\nonumber \\
	& \leq2\epsilon.
	\label{eq:U-perp}
\end{align}
 Moreover, we have already obtained
\begin{align*}
	\left\Vert \mPsi\mSigma^{\sharp}\mU^{\sharp}-\mB^{\sharp}\mU^{\sharp}\right\Vert _{F} 
	& \leq 2\epsilon.
\end{align*}
It follows from this latter bound  that $\mSigma^{\sharp}\mU^{\sharp}$ is feasible in (\ref{eq:Q-opt}). 

Since $\mPsi$ obeys \eqref{eq:rip}, we can use standard results for compressive sensing of block-sparse signals (e.g.\ \cite{eldar09ro}) to guarantee that for some absolute constant $c>0$ we have 
\begin{align*}
	\left\Vert \mQ^{\sharp}-\mSigma^{\sharp}\mU^{\sharp}\right\Vert _{F} & \leq c\epsilon.
\end{align*}
Therefore, using \eqref{eq:U-perp} and with $C=\sqrt{c^{2}+\frac{4}{1-\delta_{\sK}}}$ we have
\begin{align*}
	\left\Vert \mQ^{\sharp}\mU^{\sharp\T}\!\!-\!\!\mSigma^{\sharp}\right\Vert _{F}^{2}\! \!
	&\! =\left\Vert \mQ^{\sharp}\mU^{\sharp\T}\!\!-\!\mSigma^{\sharp}\mU^{\sharp}\mU^{\sharp\T}\right\Vert _{F}^{2} \!\!\!+\! \left\Vert \!\mSigma^{\sharp}\!\left(\!\mId\!-\!\mU^{\sharp}\mU^{\sharp\T}\!\right)\right\Vert _{F}^{2}\\
 	&\! =\left\Vert \mQ^{\sharp}-\mSigma^{\sharp}\mU^{\sharp}\right\Vert _{F}^{2}
 	+\left\Vert \mSigma^{\sharp}\left(\mId -\mU^{\sharp}\mU^{\sharp\T}\right)\right\Vert _{F}^{2}\\
 	%& \leq c^{2}\epsilon^{2}+\frac{4\epsilon^{2}}{1-\delta_{\sK}},
 	&\leq C^2\epsilon^2,
\end{align*}
and thereby 	$\left\Vert \mQ^{\sharp}\mU^{\sharp\T}-\mSigma^{\sharp}\right\Vert _{F}  \leq C\epsilon$.
%------------------------------------------------------------------------------
\section{Simulations}
\label{sec:simulations}
\begin{figure}
\noindent\centering
\includegraphics[width=1\columnwidth]{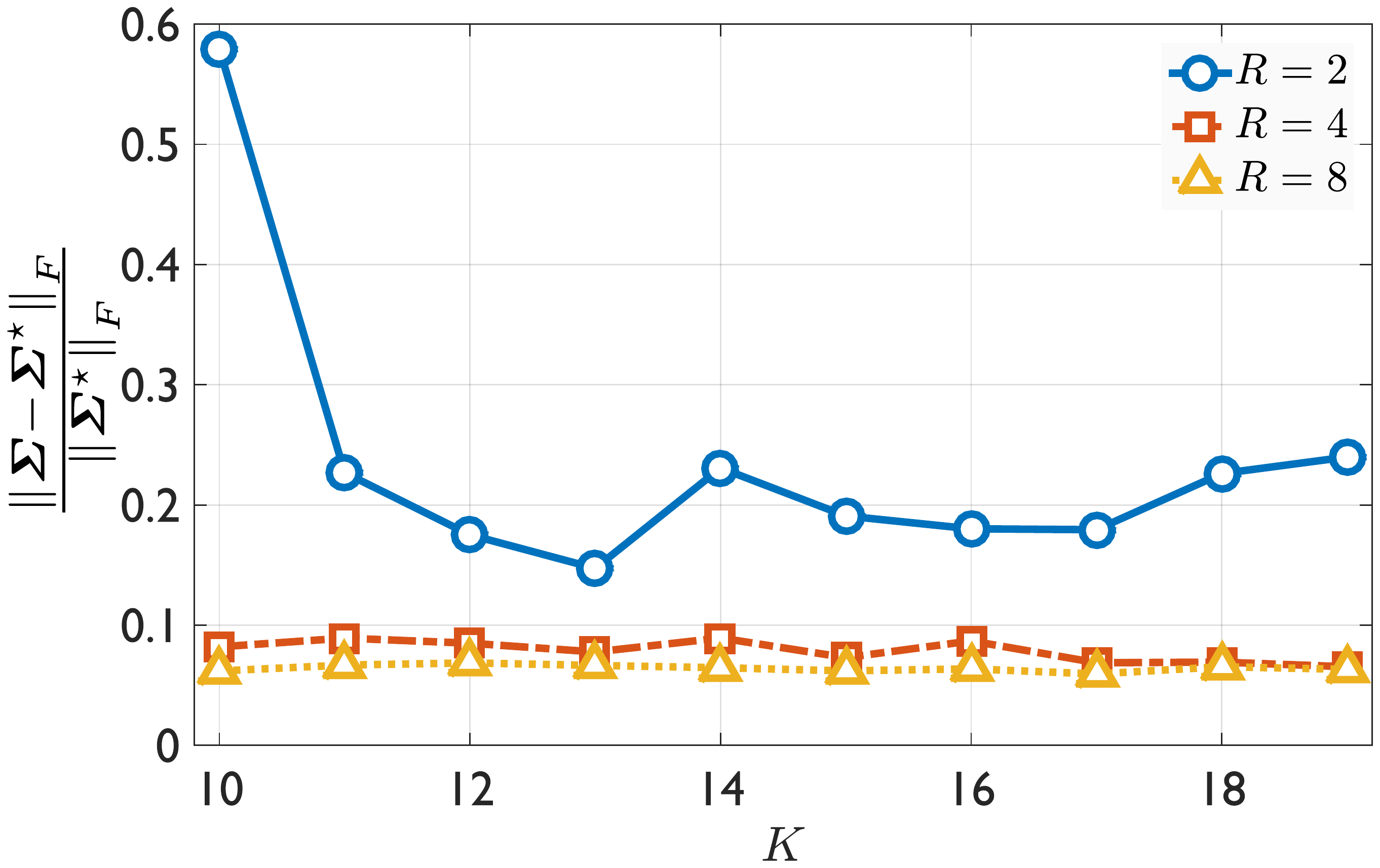}
\caption{The empirical $0.9$ quantile of estimation error vs. sparsity $\sK$ for rank $\sR$ in $\left\lbrace 2,4,8\right\rbrace$}
\label{fig:NEvsK-R}
\end{figure}
We ran numerical simulations on synthetic data as follows. With $\sN=1000$ and for $\sR\in\left\lbrace 2,4,8 \right\rbrace$ and $\sK\in\left\lbrace 10,11,\dotsc,19\right\rbrace$, in each of the $100$ trials we generated an $\sN\times \sR$ matrix $\mU$ by drawing a $\sK\times \sR$ random matrix with iid standard Gaussian entries, modulating its columns by iid $\mr{Uniform}[0,1]$, and interleaving its rows with $\sN-\sK$ all-zero rows uniformly at random. Then $\mSigma^\star = \mU\mU^\T$ is selected as the target covariance matrix. We also set $\sM=\left\lceil 2\sK\left(1+\log\frac{\sN}{\sK}\right)\right\rceil$ and $\sL=3\sR\sM$. To compute the measurements we draw $2500$ iid samples $\vx_t$ from  $\mr{Normal}\left(\mb{0},\mSigma^\star\right)$ and $\sL$ samples of $\va_\ell = \mPsi^\T\vw_\ell$ where the $\sM\times\sN$ marix $\mPsi$ is populated with iid $\mr{Normal}\left(0,\frac{1}{M}\right)$ and $\vw_\ell \sim \mr{Normal}\left(\mb{0},\mId\right)$. In each trial, we applied the proposed method with $\varepsilon = 2\sR\sK/\sqrt{\sT}$, $c_1 = \sqrt{3}$, and $c_2 =3$. Figure \ref{fig:NEvsK-R} illustrates the $0.9$ quantile of the relative error of the estimated covariance matrix as $\sK$ varies between $10$ and $19$ for $\sR=2,4$ and $8$. The error is almost flat as a function of $\sK$, in agreement with the theoretical analysis for the prescribed number of measurements. Higher variations for smaller values of $\sR$ is the effect of smaller sample size.
%------------------------------------------------------------------------------
\bibliographystyle{IEEEtran}
\bibliography{camsap-refs}

\begin{thebibliography}{10}
\providecommand{\url}[1]{#1}
\csname url@samestyle\endcsname
\providecommand{\newblock}{\relax}
\providecommand{\bibinfo}[2]{#2}
\providecommand{\BIBentrySTDinterwordspacing}{\spaceskip=0pt\relax}
\providecommand{\BIBentryALTinterwordstretchfactor}{4}
\providecommand{\BIBentryALTinterwordspacing}{\spaceskip=\fontdimen2\font plus
\BIBentryALTinterwordstretchfactor\fontdimen3\font minus
  \fontdimen4\font\relax}
\providecommand{\BIBforeignlanguage}[2]{{%
\expandafter\ifx\csname l@#1\endcsname\relax
\typeout{** WARNING: IEEEtran.bst: No hyphenation pattern has been}%
\typeout{** loaded for the language `#1'. Using the pattern for}%
\typeout{** the default language instead.}%
\else
\language=\csname l@#1\endcsname
\fi
#2}}
\providecommand{\BIBdecl}{\relax}
\BIBdecl

\bibitem{bioucas-dias14co}
J.~Bioucas-Dias, D.~Cohen, and Y.~Eldar, ``Covalsa: Covariance estimation from
  compressive measurements using alternating minimization,'' in \emph{Signal
  Processing Conference ({EUSIPCO}), 2014 Proceedings of the 22nd European},
  Sept 2014, pp. 999--1003.

\bibitem{dasarathy15sk}
G.~Dasarathy, P.~Shah, B.~Bhaskar, and R.~Nowak, ``Sketching sparse matrices,
  covariances, and graphs via tensor products,'' \emph{Information Theory,
  {IEEE} Transactions on}, vol.~61, no.~3, pp. 1373--1388, March 2015.

\bibitem{chen14es}
Y.~Chen, Y.~Chi, and A.~Goldsmith, ``Estimation of simultaneously structured
  covariance matrices from quadratic measurements,'' in \emph{Acoustics, Speech
  and Signal Processing (ICASSP), 2014 {IEEE} International Conference on}, May
  2014, pp. 7669--7673.

\bibitem{chen15ex}
------, ``Exact and stable covariance estimation from quadratic sampling via
  convex programming,'' \emph{{IEEE} Trans. Inform. Theory}, vol.~61, no.~7,
  pp. 4034 -- 4059, July 2015.

\bibitem{iwen15ro}
M.~Iwen, A.~Viswanathan, and Y.~Wang, ``Robust sparse phase retrieval made
  easy,'' \emph{Applied and Computational Harmonic Analysis}, 2015, in press.
  Preprint, \href{http://arxiv.org/abs/1410.5295}{\texttt{arXiv:1410.5295
  [math.NA]}}.

\bibitem{bahmani15ef}
S.~Bahmani and J.~Romberg, ``Efficient compressive phase retrieval with
  constrained sensing vectors,'' in \emph{the Twenty-ninth Annual Conference on
  Neural Information Processing Systems (NIPS)}, Montreal, Canada, Dec. 2015,
  preprint \href{http://arxiv.org/abs/1507.08254}{\texttt{arXiv:1507.08254
  [cs.IT]}}.

\bibitem{candes06ne}
E.~Cand\`es and T.~Tao, ``Near-optimal signal recovery from random projections:
  {U}niversal encoding strategies?'' \emph{IEEE Trans. Inform. Theory},
  vol.~52, no.~12, pp. 5406--5245, 2006.

\bibitem{vershynin12in}
R.~Vershynin, ``Introduction to the non-asymptotic theory of random matrices,''
  in \emph{Compressed Sensing, Theory and Applications}, Y.~Eldar and
  G.~Kutyniok, Eds.\hskip 1em plus 0.5em minus 0.4em\relax Cambridge University
  Press, 2012, pp. 210--268.

\bibitem{rauhut10co}
H.~Rauhut, ``Compressive sensing and structured random matrices,'' \emph{Radon
  Series Comp. Appl. Math}, vol.~9, pp. 1--92, 2010.

\bibitem{kueng14lo}
R.~Kueng, H.~Rauhut, and U.~Terstiege, ``Low rank matrix recovery from rank one
  measurements,'' \emph{Applied and Computational Harmonic Analysis}, 2015, in
  press. Preprint \href{http://arxiv.org/abs/1410.6913}{\texttt{arXiv:1410.6913
  [cs.IT]}}.

\bibitem{eldar09ro}
Y.~C. Eldar and M.~Mishali, ``Robust recovery of signals from a structured
  union of subspaces,'' \emph{IEEE Trans. Inform. Theory}, vol.~55, no.~11, pp.
  5302--5316, November 2009.

\end{thebibliography}
%\bibliography{camsap-refs-nohref}
%------------------------------------------------------------------------------
\end{document}